\documentclass[letterpaper,preprintnumbers,twocolumn,eqsecnum,superscriptaddress,aps,nofootinbib]{revtex4}
\usepackage{amssymb}\usepackage[centertags]{amsmath}\usepackage{txfonts}\usepackage{epsfig}\usepackage{bm}
\usepackage{color}\usepackage{graphicx,graphics}\usepackage{multirow}\usepackage{float}\usepackage{ulem}
\usepackage{hyperref}\usepackage{setspace}\usepackage{slashed}\usepackage{booktabs}
\hypersetup{colorlinks=true, citecolor=blue, linkcolor=red, filecolor=black,urlcolor=blue}
\allowdisplaybreaks[4]

\begin{document}

\title{Semi-inclusive deeply inelastic neutrino and antineutrino nucleus scattering}

\author{Xing-hua Yang}
\affiliation{School of Physics and Optoelectronic Engineering, Shandong University of Technology,\\ Zibo, Shandong 255000, China}

\author{Wei-hua Yang}\thanks{Corresponding author}
\affiliation{College of Nuclear Equipment and Nuclear Engineering, Yantai University,\\ Yantai, Shandong 264005, China}

\begin{abstract}
 The (anti)neutrino nucleus scattering plays a very important role in probing the hadronic structure as well as the electroweak phenomenologies. To this end, we calculate the jet production semi-inclusive deeply inelastic (anti)neutrino nucleus scattering process. The initial (anti)neutrino is assumed to be scattered off by a target particle with spin 1/2 or 1. Due to the limitation of the factorization theorem, calculations are carried out in the QCD parton model framework up to tree level twist-3. We consider both the neutral current and the charged current processes and write them into a unified form due to the similar interaction forms. Considering the angular modulations and polarizations of the cross section, we calculate the complete azimuthal asymmetries. We also calculate the intrinsic asymmetries which reveal the imbalance in the distribution of the intrinsic transverse momentum of the quark. We find that these asymmetries can be expressed in terms of the transverse momentum-dependent parton distribution functions (TMD PDFs) and the electroweak couplings. With the determined couplings, these asymmetries can be used to extract the TMD PDFs and further to study the nucleus structures.


\end{abstract}

\maketitle

\section{Introduction}

Our understanding of the parton-level structure in the nucleon comes predominantly from the lepton ($e^-,\mu^-$) deeply inelastic scattering (DIS) measurements in which parton distribution functions (PDFs) are extracted with very high precision. This process provides a great opportunity to understand the parton model and/or the factorization theorem. It will still play an important role in the future Electron-Ion Collider (EIC) ~\cite{Accardi:2012qut,AbdulKhalek:2021gbh,Anderle:2021wcy} experiment to explore the spin and three-dimensional structure of the nucleon over wide kinematic regions. In addition to the charged lepton DIS, (anti)neutrino nucleus scattering is also important in studying the nucleon and/or nucleus structures. On the one hand, it provides information on the flavor separation which cannot be realized in the charged lepton (SI)DIS experiments alone. On the other hand, the (anti)neutrino nucleus scattering can be used to study the EMC effect \cite{EuropeanMuon:1983wih} since high $Z$ nuclei are usually used in the scattering experiments because of the very weak interaction between the (anti)neutrino and the nucleus. The DIS reaction is proposed mainly for the study of strong interactions but it has abilities to study electroweak physics by considering the neutral current and charged current interactions, i.e., interactions are mediated by the $Z^0$ and $W^\pm$ bosons. For example, precision measurements of the weak mixing angle \cite{Boer:2011fh} (parity violating asymmetries \cite{Cahn:1977uu}, charge asymmetries and left-right asymmetries) can be used to determine the running of $\sin^2\theta_W$ as a function of $Q^2$, which is helpful in finding hints of new physics.
Furthermore, measurements of the neutrino oscillation, $CP$ violation, mass hierarchy and other topics rely heavily on accurate measurements of different neutrino scattering processes for different kinematic regions, such as the quasielastic scattering, inelastic scattering and the deeply inelastic scattering \cite{SajjadAthar:2020nvy,SajjadAthar:2022pjt}.

Measurable quantities in DIS are expressed in terms of PDFs which reveal the longitudinal momentum distributions of quarks and gluons or the one-dimensional structure of the nucleon. To explore the three-dimensional structures or the transverse momentum-dependent (TMD) PDFs, we need to consider the semi-inclusive DIS (SIDIS) where a final current region jet or a hadron is also measured in addition to the scattered lepton. Comparing to the hadron production SIDIS, the jet production one has two distinct features. First, the production of jets in a reaction can take on simpler forms that do not introduce additional uncertainties that arise from fragmentation functions. This is helpful to improve the measurement accuracy. However, due to the conservation of the helicity, jet production SIDIS cannot access the chiral-odd TMD PDFs. Second, the final jet in SIDIS can be a direct probe of analyzing properties of the quark transverse momentum in the $VN$ collinear frame. Here $V$ denotes the propagator vector boson ($\gamma^*, Z^0, W^\pm$). In this frame, the transverse momentum of jet is equal to that of the quark if higher order gluon radiation is neglected. Therefore the jet production SIDIS is a good candidate to explore nucleon three dimensional imaging  \cite{Song:2010pf,Song:2013sja,Wei:2016far,Gutierrez-Reyes:2018qez,Gutierrez-Reyes:2019vbx,Liu:2018trl, Liu:2020dct,Kang:2020fka,Arratia:2020ssx,H1:2021wkz}.

The electron induced jet production SIDIS for both the neutral current and the charged current cases are available in Refs. \cite{Chen:2020ugq,Yang:2020qsk}. In this paper, we calculate the (anti)neutrino induced jet production SIDIS process. It can provide not only information on the flavor separation and EMC effects, as mentioned before, but also additional ways to measure the weak mixing angle.
The initial (anti)neutrino is assumed to be scattered off by a target particle with spin 1/2 or 1. Both the neutral current and the charged current interactions are considered. Due to the limitation of the factorization theorem, we here only consider the leading order or tree level reaction. Calculations are carried out up to twist-3 level in the QCD parton model by applying the collinear expansion formalism \cite{Ellis:1982wd,Qiu:1990xxa,Liang:2006wp}. Higher twist effects are often significant for semi-inclusive reaction processes and TMD observables. For the case of twist-3 corrections, they often lead to azimuthal asymmetries which are different from the leading twist ones~\cite{Mulders:1995dh,Bacchetta:2006tn,Boer:1997mf}.
Thus, the studies of higher twist effects will give complementary or even direct access to the nucleon structure or hadronization mechanism. We therefore present the results of azimuthal asymmetries and intrinsic asymmetries after obtaining the differential cross section. The intrinsic asymmetry which was introduced in Ref. \cite{Yang:2022xwy} would reveal the imbalance in the distribution of the quark intrinsic transverse momentum. We present a rough numerical estimate to illustrate the intrinsic asymmetries. These asymmetries provide a set of new quantities for analyzing the (TMD)PDFs and the electroweak couplings in the jet production SIDIS process. It is helpful to understand the hadronic structures and the $V-A$ theory. For the charged current scattering process, we further define the $plus$ and $minus$ cross sections to study TMD PDFs.


To be explicit, we organize this paper as follows. In Sec. \ref{sec:formalism}, we present the kinematics and conventions used in the calculations. In Sec. \ref{sec:partonmodel}, we calculate the hadronic tensor in the parton model framework and show the explicit expression at twist-3 level. The differential cross sections and corresponding measurable quantities for both the neutral current and charged current cases are respectively presented in Sec. \ref{sec:neutral} and Sec. \ref{sec:charged}. A brief summary will be shown in sec. \ref{sec:summary}.

\section{The kinematics and conventions} \label{sec:formalism}

In this paper, we consider the semi-inclusive (anti)neutrino DIS in the standard model (SM) framework. Both the neutral current and charged current interactions for neutrino and antineutrino scattering are included, i.e., we calculate the following differential cross sections,
\begin{align}
  & d\sigma^{NC}_{\nu N}, && d\sigma^{NC}_{\bar\nu N}, && d\sigma^{CC}_{\nu N}, && d\sigma^{CC}_{\bar\nu N}, \label{f:cross4}
\end{align}
where $NC, CC$ denotes the neutral current and the charged current and $\nu N, \bar\nu N$ denote the neutrino nucleus scattering and antineutrino nucleus scattering.
To be explicit, we label the (anti)neutrino SIDIS as,
\begin{align}
l(l) + N(p,S) \rightarrow l^\prime (l^\prime) + q(k^\prime) + X,
\end{align}
where $l$ can be a neutrino or an antineutrino and $l^\prime$ is the corresponding final neutral or charged lepton. $N$ denotes the target particle, either a nucleon with spin $S=1/2$ or a nucleus with spin $S=1$. $q$ is used to denote the produced jet which is taken as a quark in this paper for simplicity, i.e., we do not consider the inner structure of the jet. Momenta of the incident particles are shown in the parentheses. The kinematic variables for the SIDIS are
\begin{align}
 x=\frac{Q^2}{2 p\cdot q}, \ \  y=\frac{p\cdot q}{p \cdot l},\ \ s=(p+l)^2,
\label{f:vars}
\end{align}
where $Q^2 = -q^2 =-(l-l^\prime)^2$.

Although, the neutral current and charged current in the SM have different couplings, they have the same forms. We therefore can write down the differential cross sections in a unified form,
\begin{align}
  d\sigma_r = \frac{\alpha_{\rm em}^2}{sQ^4} A_r L_{\mu\nu}(l,l^\prime)  W_r^{\mu\nu}(q,p,k^\prime) \frac{d^3 l^\prime d^3 k^\prime}{(2\pi)^3E_{l^\prime} 2E_{k^\prime}}, \label{f:crosssec}
\end{align}
where subscript $r=Z, W$ for the neutral current and charged current (anti)neutrino scattering processes, respectively. The hadronic tensor is given by
\begin{align}
  W_{r}^{\mu\nu}(q,p,k^\prime) &= \sum_X (2\pi)^3  \delta^4(p + q - k^\prime - p_X)\nonumber\\
 &\times\langle N | J_r^\mu(0)|k^\prime;X\rangle \langle k^\prime;X | J_r^\nu(0) | N \rangle,
\end{align}
where $J_r^{\mu}(0)$ is the current can be either the neutral weak current or the charged weak current. For the neutral weak current, $J_{Z}^\mu(0) =\bar\psi(0) \Gamma^\mu_q \psi(0)$ with $\Gamma^\mu_q = \gamma^\mu(c_V^q - c_A^q \gamma^5)$. For the charged weak current,  $J_{W}^\mu(0) =\bar\psi(0) \Gamma^\mu_q \psi(0)$ with $\Gamma^\mu_q = \gamma^\mu(1- \gamma^5)/2$.
It is convenient to consider the $k_\perp^\prime$-dependent hadronic tensor which is given by
\begin{align}
W_r^{\mu\nu}(q,p,k_\perp^\prime) = \int \frac{dk_z^\prime}{(2\pi)^3 2E_{k^\prime}} W_{r}^{\mu\nu}(q,p,k^\prime).
\end{align}
In terms of the variables in Eq. (\ref{f:vars}), we have $d^3 l^\prime/2E_{l^\prime}= y s dx d y d\psi/4$,
where $\psi$ is the azimuthal angle of $\vec l^\prime$ around $\vec l$.
Therefore the cross section can be rewritten as
\begin{align}
\frac{d\sigma_r}{dx dy d\psi d^2 k_\perp^\prime} = \frac{y \alpha_{\rm em}^2}{2 Q^4} A_{r} L_{\mu\nu}^r(l, l^\prime) W_r^{\mu\nu}(q,p,k_\perp^\prime).
\label{f:crossfinal}
\end{align}

The propagator factor for the neutral current and charged current are respectively given by
\begin{align}
& A_{Z} = \frac{Q^4}{\left[(Q^2+M_Z^2)^2 + \Gamma_Z^2 M_Z^2 \right] \sin^4 2\theta_W}, \\
& A_{W} = \frac{Q^4}{\left[(Q^2+M_W^2)^2 + \Gamma_W^2 M_W^2 \right] 16\sin^4 \theta_W},
\end{align}
where $\Gamma_{Z,W}$ and $M_{Z,W}$ are widths and masses for $Z^0$ boson and $W^\pm$ boson, respectively. $\theta_W$ is the weak mixing angle. The leptonic tensor is defined as
\begin{align}
 &L_{\mu\nu}(l,l^\prime)= 2\left[ l_\mu l^\prime_\nu + l_\nu l^\prime_\mu - (l\cdot l^\prime)g_{\mu\nu}  \right] \pm 2i\lambda_\nu \varepsilon_{\mu\nu l l^\prime}. \label{f:leptongamma}
\end{align}
where $\lambda_\nu$ is introduced for convenience. For neutrino the sign is $-$, while for antineutrino the sign is $+$.

\section{The hadronic tensor in the QCD parton model} \label{sec:partonmodel}

The hadronic tensor given in the previous section contains nonperturbative information and thus cannot be calculated with perturbative theory. To obtain the cross section, we would decompose the hadronic tensor with Lorentz invariants and subsequently write it down in terms of the structure functions which are similar to that, e.g., $F_{2}(x, Q^2)$, in the inclusive DIS process. Supposing that the target particle is a spin $S=1$ one, we obtain in total 81 structure functions in the jet production SIDIS process, see Refs. \cite{Chen:2020ugq,Yang:2020qsk}. However, we do not present them in this paper for simplicity and only concentrate on the calculations in the QCD parton model.

\begin{figure}
\centering
\includegraphics[width=1\linewidth]{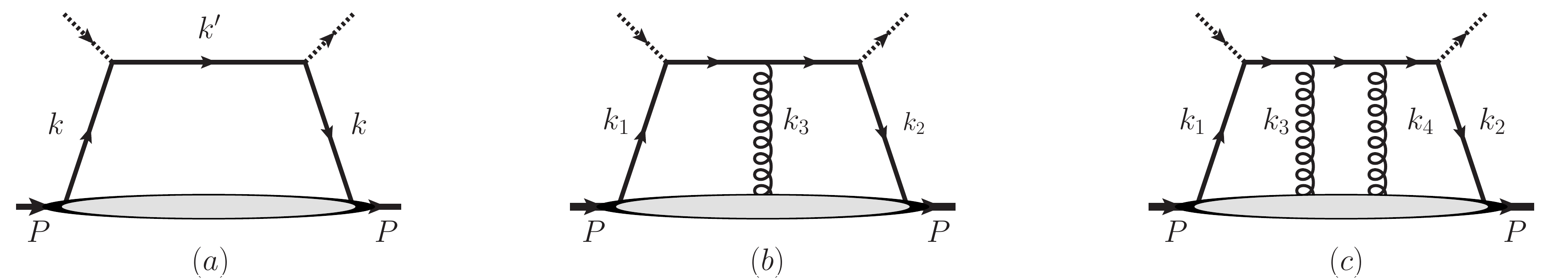}
\caption{The first few diagrams of the Feynman diagram series with exchange of $j$ gluons, where $j=0,~1$ and $2$ for diagrams $(a)$, $(b)$ and $(c)$ respectively. $(a)$ denotes the handbag diagram. $(b)$ and $(c)$ denote the multiple gluon scattering diagrams. The dashed lines can be $Z^0$ boson or the $W^\pm$ boson for the neutral and charged current interactions, respectively.}
\label{fig:neut}
\end{figure}

In the QCD parton model, the hadronic tensor is expressed by the gauge-invariant TMD PDFs in the SIDIS process. At the tree level, we need to consider the contributions from the series of diagrams shown in Fig.~\ref{fig:neut}. The dashed lines denote $Z^0$ boson or the $W^\pm$ boson for the neutral and charged current interactions, respectively. One only needs to consider the handbag diagram, (a), for calculating the leading twist (twist-2) hadronic tensor. For higher twist ones, multiple gluon scattering diagrams, (b) and (c), should be included.

The hadronic tensor is spin-dependence. To be explicit, we parametrize polarizations in the following. For the spin 1/2 particle, the spin vector can be parametrized as $ S^\mu =\lambda_h \frac{p^+}{M}\bar n^\mu + S_T^\mu -\lambda_h \frac{M}{2p^+}n^\mu$, where $\lambda_h$ is the helicity and $M$ is the mass. The transverse vector polarization is parametrized as
\begin{align}
& S_T^\mu = |S_T| \left( 0,0, \cos\varphi_S, \sin\varphi_S \right).
\end{align}
For the spin 1 particle, there are five kinds of polarizations, $S_{LL}, S_{LT}^x, S_{LT}^y, S_{TT}^{xx}$ and $S_{TT}^{xy}$. We parametrize and define them as in Ref. \cite{Bacchetta:2000jk}, i.e.,
\begin{align}
& S_{LT}^x = |S_{LT}| \cos\varphi_{LT}, \\
& S_{LT}^y = |S_{LT}| \sin\varphi_{LT}, \\
& S_{TT}^{xx} = -S_{TT}^{yy} = |S_{TT}| \cos2\varphi_{TT}, \\
& S_{TT}^{xy} = S_{TT}^{yx} = |S_{TT}| \sin2\varphi_{TT}.
\end{align}

We note that in this section we only calculate the hadronic tensor for the neutral current process and the subscript $Z$ is neglected. For the charged current process it can be obtained similarly, we do not repeat here for simplicity.

At the leading order of the QCD, the hadronic tensor at twist-3 can be written as
\begin{align}
W_{\mu\nu} (q,p,k^\prime_\perp) = \sum_{j,c} \tilde W_{r,\mu\nu}^{(j,c)} (q,p,k^\prime_\perp),
\end{align}
after applying the collinear expansion formalism \cite{Liang:2006wp}. Here $j$ denotes the number of exchanging gluons and $c$ denotes different cuts, $c=R, L$. After simple algebraic calculations, the hadronic tensor for $0, 1$ exchanging gluons are respectively  given by
\begin{align}
& \tilde{W}_{\mu\nu}^{(0)}(q,p,k_\perp^\prime) = \frac{1}{2}{\rm Tr}\left[\hat{h}_{r,\mu \nu}^{(0)} \hat{\Phi}^{(0)}(x,k_\perp)\right] , \label{f:W0munu}\\
& \tilde{W}_{\mu\nu}^{(1, L)}(q,p,k_\perp^\prime) = \frac{1}{4 p \cdot q}{\rm Tr}\left[\hat{h}_{r,\mu\nu}^{(1) \rho} \hat{\Phi}_{\rho}^{(1)}(x,k_\perp)\right], \label{f:W1Lmunu}
\end{align}
where $\hat h$'s denote the hard parts or partonic scattering amplitudes and they are written as
\begin{align}
& \hat h_{\mu\nu}^{(0)} = \frac{1}{p^+}\Gamma_{\mu}^q \slashed n \Gamma_{\nu}^q , \qquad \hat{h}_{\mu\nu}^{(1) \rho}=\Gamma_{\mu}^q \slashed{\bar n} \gamma_{\perp}^{\rho} \slashed n \Gamma_{\nu}^q.
\end{align}
As mentioned in the previous section, $\Gamma_{\mu}^q$ are different for the neutral current and the charged current interactions. But the correlators are not affected because they are QCD quantities.
The gauge-invariant quark-quark and quark-gluon-quark correlators are defined as
\begin{align}
  \hat{\Phi}^{(0)}\left(x, k_{\perp}\right) =& \int \frac{p^{+} d y^{-} d^{2} y_{\perp}}{(2 \pi)^{3}} e^{i x p^{+} y^{-}-i \vec{k}_{\perp} \cdot \vec{y}_{\perp}} \nonumber\\
  &\times \langle N|\bar{\psi}(0) {\cal L}(0, y) \psi(y)| N\rangle, \\
  \hat{\Phi}_{\rho}^{(1)}\left(x, k_{\perp}\right) =& \int \frac{p^{+} d y^{-} d^{2} y_{\perp}}{(2 \pi)^{3}} e^{i x p^+ y^- - i \vec{k}_\perp \cdot \vec{y}_\perp}\nonumber\\
  &\times  \langle N | \bar{\psi}(0) D_{\perp \rho}(0) {\cal L}(0, y) \psi(y)| N\rangle,
\end{align}
where $D_\rho(y) = -i\partial_\rho + g A_\rho(y)$ is the covariant derivative.
${\cal L}(0, y)$ is the gauge link obtained from the collinear expansion procedure, which guarantees the gauge invariance of the correlators.

The quark-quark and quark-gluon-quark correlators can be decomposed in terms of the Dirac gamma matrices and corresponding coefficient functions because they are $4\times 4$ matrices. The complete decomposition of the correlator for the spin 1 particle can be found in Refs. \cite{Bacchetta:2000jk,Kumano:2020ijt}.
In the jet production SIDIS process only the chiral-even TMD PDFs are involved.
Following the conventions in Refs. \cite{Chen:2020ugq,Yang:2020qsk}, we decompose the correlators as $\hat \Phi= \frac{1}{2}\left[\gamma^\alpha \Phi_\alpha + \gamma^\alpha\gamma_5 \tilde\Phi_\alpha \right]$. The TMD PDFs are obtained by decomposing the coefficient functions. For $\Phi^{(0)}_\alpha$ and $\tilde\Phi^{(0)}_\alpha$, we have
\begin{align}
  \Phi^{(0)}_\alpha &=p^+ \bar n_\alpha\Bigl(f_1+S_{LL}f_{1LL}-\frac{k_\perp \cdot \tilde S_T}{M}f^\perp_{1T} \nonumber\\ &+\frac{k_\perp \cdot S_{LT}}{M}f_{1LT}^{\perp}+\frac{S_{TT}^{kk}}{M^2}f_{1TT}^{\perp} \Bigr) +k_{\perp\alpha}\Big( f^\perp +S_{LL}f_{LL}^\perp  \Big)  \nonumber\\
  &- M\tilde S_{T\alpha}f_T + M S_{LT\alpha}f_{LT}+ S^k_{TT\alpha} f_{TT} - \lambda_h \tilde k_{\perp\alpha} f_L^\perp \nonumber\\ & -\frac{k_{\perp\langle\alpha}k_{\perp\beta\rangle}}{M} \Bigl( \tilde S_T^\beta f_T^\perp + S_{LT}^\beta f_{LT}^\perp + \frac{ S_{TT}^{k\beta}}{M} f_{TT}^\perp \Bigr),
\label{eq:Xi0Peven}\\
  \tilde\Phi^{(0)}_\alpha &=p^+\bar n_\alpha\Bigl(-\lambda_hg_{1L}+\frac{k_\perp\cdot S_T}{M}g^\perp_{1T}\nonumber\\
  &+ \frac{k_\perp \cdot \tilde S_{LT}}{M} g_{1LT}^\perp - \frac{\tilde S_{TT}^{k k}}{M^2} g_{1TT}^\perp\Bigr)-\tilde k_{\perp\alpha} \Big( g^\perp+S_{LL}g_{LL}^\perp \Big) \nonumber\\
  &- M S_{T\alpha}g_T - M \tilde S_{LT\alpha}g_{LT} - \tilde S^{k}_{TT\alpha} g_{TT} -\lambda_h k_{\perp\alpha} g_L^\perp \nonumber\\
  &+ \frac{k_{\perp\langle\alpha}k_{\perp\beta\rangle}}{M} \Bigl( S_T^\beta g_T^\perp - \tilde S_{LT}^\beta g_{LT}^\perp - \frac{ S_{TT}^{k\beta}}{M} \tilde g_{TT}^\perp \Bigr).
\label{eq:Xi0Podd}
\end{align}
For $\Phi_{\rho\alpha}^{(1)}$ and $\tilde\Phi_{\rho\alpha}^{(1)}$, we have
\begin{align}
  \Phi^{(1)}_{\rho\alpha}&=p^+\bar n_\alpha\Biggl[k_{\perp\rho}\big( f^\perp_d+ S_{LL}f_{dLL}^\perp\big)- M\tilde S_{T\rho}f_{dT} \nonumber\\
  & +MS_{LT\rho} f_{dLT}+S_{TT\rho}^k f_{dTT} -\lambda_h \tilde k_{\perp\rho} f_{dL}^\perp \nonumber\\
  & -\frac{k_{\perp\langle\rho}k_{\perp\beta\rangle}}{M} \Bigl( \tilde S_T^\beta f_{dT}^\perp + S_{LT}^\beta f_{dLT}^\perp + \frac{S_{TT}^{k\beta}}{M} f_{dTT}^\perp \Bigr)\Biggr], \label{eq:Xi1Peven} \\
  \tilde \Phi^{(1)}_{\rho\alpha}&=ip^+\bar n_\alpha\Biggl[\tilde k_{\perp\rho}\big( g^\perp_d+ S_{LL}g_{dLL}^\perp\big) + MS_{T\rho}g_{dT}\nonumber\\
  & + M\tilde S_{LT\rho} g_{dLT} + \tilde S_{TT\rho}^{k} g_{dTT}+\lambda_h k_{\perp\rho} g_{dL}^\perp  \nonumber\\
  & - \frac{k_{\perp\langle\rho}k_{\perp\beta\rangle}}{M} \Bigl( S_T^\beta g_{dT}^\perp - \tilde S_{LT}^\beta g_{dLT}^\perp - \frac{\tilde S_{TT}^{k\beta}}{M} g_{dTT}^\perp \Bigr) \Biggr],
\label{eq:Xi1Podd}
\end{align}
where
$S_{TT}^{kk} \equiv S_{TT}^{\mu\nu} k_{\perp\mu} k_{\perp\nu}$,
$S_{TT}^{k\beta} \equiv S_{TT}^{\alpha\beta}k_{\perp\alpha}$, and $\tilde S_{TT}^{k\beta} \equiv \varepsilon_{\perp\mu}^{\beta} S_{TT}^{k\mu}$. Subscript $d$ denotes that the TMD PDFs are defined via the quark-gluon-quark correlator.

The leading twist and twist-3 hadronic tensors are respectively given by \cite{Chen:2020ugq,Yang:2020qsk},
\begin{widetext}
\begin{align}
\tilde W_{t2}^{\mu\nu} =&
-\left( c_1^q g_{\perp}^{\mu\nu} + ic_3^q \varepsilon_{\perp}^{\mu\nu} \right) \Bigl(f_1+S_{LL}f_{1LL}-\frac{k_\perp \cdot \tilde S_T}{M}f^\perp_{1T}+\frac{k_\perp \cdot S_{LT}}{M}f_{1LT}^{\perp}+\frac{S_{TT}^{kk}}{M^2}f_{1TT}^{\perp} \Bigr) \nonumber\\
& - \left( c_3^q g_{\perp}^{\mu\nu} + ic_1^q \varepsilon_{\perp}^{\mu\nu} \right) \Bigl(-\lambda_hg_{1L}+\frac{k_\perp\cdot S_T}{M}g^\perp_{1T} + \frac{k_\perp \cdot \tilde S_{LT}}{M} g_{1LT}^\perp - \frac{\tilde S_{TT}^{k k}}{M^2} g_{1TT}^\perp\Bigr). \label{f:Wt2leadingmunu}
\end{align}

\begin{align}
(p \cdot q) \tilde W_{t3}^{\mu\nu}
&= \Bigl[ c_1^q k_\perp^{\{\mu} \bar q^{\nu\}} + ic_3^q \tilde k_\perp^{[\mu} \bar q^{\nu]} \Bigr] \left( f^\perp + S_{LL} f_{LL}^\perp \right)- \Bigl[ c_1^q \tilde k_\perp^{\{\mu} \bar q^{\nu\}} - ic_3^q k_\perp^{[\mu} \bar q^{\nu]} \Bigr] \lambda_h f_L^\perp - \Bigl[ c_1^q \tilde S_T^{\{\mu} \bar q^{\nu\}} - ic_3^q S_T^{[\mu} \bar q^{\nu]} \Bigr] M f_T \nonumber\\
&+ \Bigl[ c_1^q S_{LT}^{\{\mu} \bar q^{\nu\}} + ic_3^q \tilde S_{LT}^{[\mu} \bar q^{\nu]} \Bigr] M f_{LT} + \Bigl[ c_1^q S_{TT}^{k\{\mu} \bar q^{\nu\}} + ic_3^q \tilde S_{TT}^{k[\mu} \bar q^{\nu]} \Bigr] f_{TT} \nonumber\\
&- \Biggl[ c_1^q \left( \frac{k_\perp\cdot \tilde S_T}{M} k_\perp^{\{\mu} \bar q^{\nu\}} - \frac{k_\perp^2}{2M} \tilde S_T^{\{\mu} \bar q^{\nu\}} \right) + ic_3^q \left(\frac{k_\perp\cdot S_T}{M} k_\perp^{[\mu} \bar q^{\nu]} - \frac{k_\perp^2}{2M}S_T^{[\mu} \bar q^{\nu]} \right) \Biggr] f_T^\perp \nonumber\\
&- \Biggl[ c_1^q \left( \frac{k_\perp\cdot S_{LT}}{M} k_\perp^{\{\mu} \bar q^{\nu\}}- \frac{k_\perp^2}{2M} S_{LT}^{\{\mu} \bar q^{\nu\}} \right) + ic_3^q \left(\frac{k_\perp\cdot S_{LT}}{M}\tilde k_\perp^{[\mu} \bar q^{\nu]} - \frac{k_\perp^2}{2M}\tilde S_{LT}^{[\mu} \bar q^{\nu]} \right) \Biggr] f_{LT}^\perp \nonumber\\
&- \Biggl[ c_1^q \left( \frac{S_{TT}^{kk}}{M^2} k_\perp^{\{\mu} \bar q^{\nu\}} - \frac{k_\perp^2}{2M^2} S_{TT}^{k\{\mu} \bar q^{\nu\}} \right) + ic_3^q \left(\frac{S_{TT}^{kk}}{M^2}\tilde k_\perp^{[\mu} \bar q^{\nu]} - \frac{k_\perp^2}{2M^2}\tilde S_{TT}^{k[\mu} \bar q^{\nu]} \right) \Biggr] f_{TT}^\perp \nonumber\\
&- \Bigl[ c_3^q \tilde k_\perp^{\{\mu} \bar q^{\nu\}} - ic_1^q k_\perp^{[\mu} \bar q^{\nu]}\Bigr] \left( g^\perp + S_{LL} g_{LL}^\perp \right)- \Bigl[ c_3^q k_\perp^{\{\mu} \bar q^{\nu\}} + ic_1^q \tilde k_\perp^{[\mu} \bar q^{\nu]}\Bigr] \lambda_h g_L^\perp - \Bigl[ c_3^q S_T^{\{\mu} \bar q^{\nu\}} + ic_1^q \tilde S_T^{[\mu} \bar q^{\nu]} \Bigr] M g_T \nonumber\\
&- \Bigl[ c_3^q \tilde S_{LT}^{\{\mu} \bar q^{\nu\}} - ic_1^q S_{LT}^{[\mu} \bar q^{\nu]} \Bigr] M g_{LT} - \Bigl[ c_3^q \tilde S_{TT}^{k\{\mu} \bar q^{\nu\}} - ic_1^q S_{TT}^{k[\mu} \bar q^{\nu]} \Bigr] g_{TT} \nonumber\\
&+ \Biggl[ c_3^q \left( \frac{k_\perp\cdot S_T}{M} k_\perp^{\{\mu} \bar q^{\nu\}} - \frac{k_\perp^2}{2M} S_T^{\{\mu} \bar q^{\nu\}} \right) + ic_1^q \left(\frac{k_\perp\cdot S_T}{M} \tilde k_\perp^{[\mu} \bar q^{\nu]} - \frac{k_\perp^2}{2M} \tilde S_T^{[\mu} \bar q^{\nu]} \right) \Biggr] g_T^\perp \nonumber\\
&- \Biggl[ c_3^q \left( \frac{k_\perp\cdot \tilde S_{LT}}{M} k_\perp^{\{\mu} \bar q^{\nu\}} - \frac{k_\perp^2}{2M} \tilde S_{LT}^{\{\mu} \bar q^{\nu\}} \right) + ic_1^q \left(\frac{k_\perp\cdot S_{LT}}{M} k_\perp^{[\mu} \bar q^{\nu]} - \frac{k_\perp^2}{2M} S_{LT}^{[\mu} \bar q^{\nu]} \right) \Biggr] g_{LT}^\perp \nonumber\\
&- \Biggl[ c_3^q \left( \frac{k_\perp\cdot \tilde S_{TT}^{k}}{M^2} k_\perp^{\{\mu} \bar q^{\nu\}}- \frac{k_\perp^2}{2M^2} \tilde S_{TT}^{k\{\mu} \bar q^{\nu\}} \right) + ic_1^q \left(\frac{k_\perp\cdot S_{TT}^{k}}{M^2} k_\perp^{[\mu} \bar q^{\nu]} - \frac{k_\perp^2}{2M^2} S_{TT}^{k[\mu} \bar q^{\nu]} \right) \Biggr] g_{TT}^\perp, \label{f:Wt3munu}
\end{align}
\end{widetext}
where $\bar q^\mu = q^\mu + 2xp^\mu$.
From the equalities, $q\cdot\bar q = q\cdot k_\perp = 0$ and $q\cdot S_T = q\cdot S_{LT} = q\cdot S_{TT}^{k}/M = 0$, we see clearly that the full twist-3 hadronic tensor satisfies current conservation law, $q_\mu \tilde W^{\mu\nu}_{t3} = q_\nu \tilde W^{\mu\nu}_{t3} = 0$.

We emphasize that the hadronic tensor for the neutral current and the charged current interactions is different. Since the correlators are QCD quantities and remain unchanged. Therefore the difference lies in the hart parts $\hat h_r$'s. To be more precise, the difference is the weak couplings, $c_V^q, c_A^q$. For charged current interaction process, according to our definition, $c_V^q= c_A^q =1$. Therefore, we can only calculate the hadronic tensor or the following differential cross section for the neutral current interaction and obtain the charged current one by setting $c_V^q= c_A^q =1$.

\section{Results of the neutral current process} \label{sec:neutral}

In expressing the cross section, we choose the $V N$ collinear frame, in which momenta related to this neutrino SIDIS process take the following forms:
\begin{align}
& p^\mu = \left(p^+,0,\vec 0_\perp \right), \nonumber\\
& l^\mu = \left( \frac{1-y}{y}xp^+, \frac{Q^2}{2xyp^+}, \frac{Q\sqrt{1-y}}{y},0 \right),\nonumber\\
& q^\mu = \left( -xp^+, \frac{Q^2}{2xp^+}, \vec 0_\perp \right), \nonumber\\
& k_\perp^{\prime\mu} = k_\perp^\mu = |k_\perp| \left( 0,0, \cos\varphi, \sin\varphi \right).
\end{align}
In this frame, the transverse momenta of the jet ($k_\perp^\prime$) and that of the quark ($k_\perp$) inside the nucleon are the same. We do not distinguish them in the following calculations. This is also the precondition that we calculate the intrinsic asymmetries in the following context.



\subsection{The differential cross section}

Similar to dealing with the hadronic tensor, we divide the cross section into a leading twist part and a twist-3 part.
Substituting the leading twist hadronic tensor and the leptonic tensor into Eq.~(\ref{f:crossfinal}) yields the leading twist cross section.
It is given by
\begin{align}
  \frac{d\sigma_{\nu N,t2}^{NC}}{dx dy d\psi d^2 k_\perp^\prime} =\frac{\alpha_{\rm em}^2 A_{Z}}{y Q^2}&\Biggl\{ T_0^q(y)(f_1+S_{LL}f_{1LL})\nonumber \\
   &- T_1^q(y)\lambda_h g_{1L} \nonumber\\
  +|S_T|k_{\perp M}&\Big[\sin(\varphi-\varphi_S)T_0^q(y) f^\perp_{1T}\nonumber\\
  -&\cos(\varphi-\varphi_S) T_1^q(y)g^\perp_{1T}\Big] \nonumber\\
  -|S_{LT}|k_{\perp M}&\Big[\sin(\varphi-\varphi_{LT})T_1^q(y) g^\perp_{1LT}\nonumber\\
  +&\cos(\varphi-\varphi_{LT}) T_0^q(y) f^\perp_{1LT}\Big] \nonumber\\
  -|S_{TT}|k_{\perp M}^2&\Big[\sin(2\varphi-2\varphi_{TT})T_1^q(y) g^\perp_{1TT}\nonumber\\
  - &\cos(2\varphi-2\varphi_{TT})T_0^q(y)f^\perp_{1TT}\Big]
 \Biggr\}, \label{f:crosst2NCnu}
\end{align}
where subscript $t2$ denotes leading twist. Here we have defined $k_{\perp M} = |k_\perp|/M$, and
\begin{align}
  & T_0^q(y) = c_1^q A(y) + c_3^q C(y), \label{f:T0y} \\
  & T_1^q(y) = c_3^q A(y) + c_1^q C(y), \label{f:T1y}
\end{align}
to simplify the expression. Here $A(y) = y^2-2y+2, C(y) = y(2-y)$.


Substituting the twist-3 hadronic tensor and the leptonic tensor into Eq. (\ref{f:crossfinal}) yields the twist-3 cross section. It is given by
\begin{align}
  \frac{d\sigma_{\nu N,t3}^{NC}}{dx dy d\psi d^2 k_\perp^\prime}& = \frac{\alpha_{\rm{em}}^2 A_{Z} }{-y Q^2}2x\kappa_M \Biggl\{ \lambda_h k_{\perp M}\nonumber\\
  &\times\Big[\sin\varphi T_2^q(y)f^\perp_L- \cos\varphi T_3^q(y)g_L^\perp\Big] \nonumber\\
  &+k_{\perp M}\cos\varphi T_2^q(y)(f^\perp+S_{LL}f^\perp_{LL}) \nonumber\\
  &+k_{\perp M}\sin\varphi  T_3^q(y)(g^\perp+S_{LL}g^\perp_{LL}) \nonumber\\
  &+|S_T|\Big[\sin\varphi_S  T_2^q(y)f_T -\cos\varphi_S T_3^q(y)g_T  \nonumber\\
  &\quad \quad +\sin(2\varphi-\varphi_S) T_2^q(y)\frac{k_{\perp M}^2}{2}f^\perp_T \nonumber\\
  &\quad \quad -\cos(2\varphi-\varphi_S) T_3^q(y)\frac{k_{\perp M}^2}{2}g^\perp_T \Big] \nonumber\\
  &+|S_{LT}|\Big[\sin\varphi_{LT} T_3^q(y)g_{LT} +\cos\varphi_{LT} T_2^q(y)f_{LT}  \nonumber\\
  &\quad \quad +\sin(2\varphi-\varphi_{LT})T_3^q(y)\frac{k_{\perp M}^2}{2} g^\perp_{LT} \nonumber\\
  &\quad \quad +\cos(2\varphi-\varphi_{LT})T_2^q(y)\frac{k_{\perp M}^2}{2} f^\perp_{LT} \Big] \nonumber\\
  &+|S_{TT}|\Big[\sin(\varphi-2\varphi_{TT}) T_3^q(y)k_{\perp M} g_{TT} \nonumber\\
  &\quad \quad -\cos(\varphi-2\varphi_{TT}) T_2^q(y)k_{\perp M} f_{TT} \nonumber\\
  &\quad -\sin(3\varphi-2\varphi_{TT}) T_3^q(y)\frac{k_{\perp M}^3}{2} g^\perp_{TT} \nonumber\\
  &\quad -\cos(3\varphi-2\varphi_{TT}) T_2^q(y)\frac{k_{\perp M}^3}{2} f^\perp_{TT} \Big]
 \Biggr\}, \label{f:crosst3NCnu}
\end{align}
where subscript $t3$ denotes twist-3. The twist suppression factor is defined as $\kappa_M=M/Q$.
We have also defined
\begin{align}
  & T_2^q(y) = c_1^q B(y) + c_3^q D(y), \\
  & T_3^q(y) = c_3^q B(y) + c_1^q D(y),
\label{eq:T2T3}
\end{align}
where $B(y) = 2(2-y)\sqrt{1-y}, D(y) = 2y\sqrt{1-y}$.

Equations (\ref{f:crosst2NCnu}) and (\ref{f:crosst3NCnu}) give the differential cross section of the neutral current neutrino nucleus scattering. To obtain the cross section corresponding to the antineutrino nucleus scattering, we only need to replace $T^q_{0,1,2,3}(y)$ by  $t^q_{0,1,2,3}(y)$ which are defined as
\begin{align}
  & t_0^q(y) = c_1^q A(y) - c_3^q C(y), \\
  & t_1^q(y) = c_3^q A(y) - c_1^q C(y), \\
  & t_2^q(y) = c_1^q B(y) - c_3^q D(y), \\
  & t_3^q(y) = c_3^q B(y) - c_1^q D(y).
\end{align}

\subsection{The azimuthal asymmetries}

One of the important measurable quantities in the high energy nucleon reaction is the azimuthal asymmetry. Azimuthal asymmetries can be measured to extract TMD PDFs and further to study the three dimensional imaging of the nucleon. In this subsection, we calculate the azimuthal asymmetries induced by the final jet. We first introduce the definition
\begin{align}
  \langle \cos\varphi \rangle_{U,U}=\frac{\int d\tilde\sigma \cos\varphi d\varphi}{\int d\tilde\sigma d\varphi},
\end{align}
for the unpolarized and the longitudinally polarized target case, and
\begin{align}
  \langle \cos(\varphi-\varphi_S) \rangle_{U,T}=\frac{\int d\tilde\sigma \cos(\varphi-\varphi_S)d\varphi d\varphi_S}{\int d\tilde\sigma d\varphi d\varphi_S},
\end{align}
for the transversely polarized target case.
$d\tilde\sigma$ is used to denote $\frac{d\sigma}{dx dy d\psi d^2 k_\perp^\prime}$,
and $d\varphi_S\approx d\psi$ when the direction is chosen to be the direction of $S$ in case of a transversely polarized target in which integration corresponds to take the average over the out going electron's azimuthal angle~\cite{Bacchetta:2006tn,Diehl:2005pc}.
The subscripts such as $U, T$ denote the polarizations of the lepton beam and the target, respectively.

At the leading twist, there are 6 transversely polarized dependent azimuthal asymmetries which are given by
\begin{align}
 & \langle \sin(\varphi-\varphi_S) \rangle_{U,T} = k_{\perp M} \frac{T_{0}^q(y) f^\perp_{1T}}{2 T_{0}^q(y) f_1}, \\
 & \langle \cos(\varphi-\varphi_S) \rangle_{U,T} = - k_{\perp M}\frac{T_{1}^q(y) g^\perp_{1T}}{2 T_{0}^q(y) f_1}, \\
 & \langle \sin(\varphi-\varphi_{LT})\rangle_{U,LT} =-k_{\perp M}\frac{T_{1}^q(y) g^\perp_{1LT}}{2 T_{0}^q(y)f_1}, \\
 & \langle \cos(\varphi-\varphi_{LT})\rangle_{U,LT} =-k_{\perp M}\frac{T_{0}^q(y) f^\perp_{1LT}}{2 T_{0}^q(y)f_1}, \\
 & \langle \sin(2\varphi-\varphi_{TT}) \rangle_{U,TT} = - k_{\perp M}^2 \frac{ T_{1}^q(y)g^\perp_{1TT}}{2 T_{0}^q(y)f_1}, \\
 & \langle \cos(2\varphi-\varphi_{TT}) \rangle_{U,TT} = k_{\perp M}^2 \frac{ T_{0}^q(y)f^\perp_{1TT}}{2 T_{0}^q(y)f_1}.
\end{align}
In addition to the TMD PDFs, azimuthal asymmetries also depends on $T$ functions shown in Eqs. (\ref{f:T0y}) and (\ref{f:T1y}). This implies that TMD PDFs and weak couplings can be determined simultaneously by global fit.

Higher twist effects often lead to azimuthal asymmetries which are different from the leading twist ones. In the jet production SIDIS process,  we obtain 18 azimuthal asymmetries at twist-3, which are given by
\begin{align}
  & \langle \cos\varphi \rangle_{U,U} = -x\kappa_M k_{\perp M} \frac{ T_{2}^q(y)f^\perp}{T_{0}^q(y)f_1}, \\
  & \langle \sin\varphi \rangle_{U,U} = -x\kappa_M k_{\perp M} \frac{T_{3 }^q(y)g^\perp}{T_{0 }^q(y)f_1}, \\
  & \langle \cos\varphi \rangle_{U,L} = -x\kappa_M k_{\perp M} \frac{T_{2}^q(y) f^\perp -\lambda_h T_{3}^q(y)g^\perp_L}{T_{0}^q(y) f_1}, \\
  & \langle \sin\varphi \rangle_{U,L} = -x\kappa_M k_{\perp M} \frac{T_{3}^q(y) g^\perp + \lambda_h T_{2}^q(y)f^\perp_L}{T_{0}^q(y)f_1}, \\
  & \langle \cos\varphi \rangle_{U,LL} = -x\kappa_M k_{\perp M} \frac{T_{2 }^q(y)(f^\perp + S_{LL}f^\perp_{LL})}{T_{0}^q(y)f_1}, \\
  & \langle \sin\varphi \rangle_{U,LL} = -x\kappa_M k_{\perp M} \frac{T_{3 }^q(y) (g^\perp + S_{LL}g^\perp_{LL})}{T_{0}^q(y)f_1}, \\
  & \langle \cos\varphi_S \rangle_{U,T} =  x\kappa_M\frac{T_{3 }^q(y)g_T}{T_{0 }^q(y)f_1}, \\
  & \langle \sin\varphi_S \rangle_{U,T} = -x\kappa_M\frac{T_{2 }^q(y)f_T}{T_{0 }^q(y)f_1}, \\
  & \langle \cos(2\varphi-\varphi_S) \rangle_{U,T} = x\kappa_M k_{\perp M}^2\frac{ T_{3}^q(y)g^\perp_T}{2T_{0 }^q(y)f_1}, \\
  & \langle \sin(2\varphi-\varphi_S) \rangle_{U,T} = -x\kappa_M k_{\perp M}^2\frac{T_{2 }^q(y)f^\perp_T}{2T_{0 }^q(y)f_1}, \\
  & \langle \cos\varphi_{LT} \rangle_{U,LT} = - x\kappa_M\frac{T_{2}^q(y)f_{LT}}{T_{0 }^q(y)f_1}, \\
  & \langle \sin\varphi_{LT} \rangle_{U,LT} = -x\kappa_M\frac{T_{3}^q(y)g_{LT}}{T_{0 }^q(y)f_1}, \\
  & \langle \cos(2\varphi-\varphi_{LT}) \rangle_{U,LT} = -x\kappa_M k_{\perp M}^2\frac{ T_{2}^q(y)f^\perp_{LT}}{2 T_{0 }^q(y)f_1}, \\
  & \langle \sin(2\varphi-\varphi_{LT}) \rangle_{U,LT} = -x\kappa_M k_{\perp M}^2\frac{ T_{3}^q(y)g^\perp_{LT}}{2 T_{0 }^q(y)f_1}, \\
  & \langle \cos(\varphi-2\varphi_{TT}) \rangle_{U,TT} =  x\kappa_M k_{\perp M} \frac{  T_{2}^q(y)f_{TT}}{T_{0 }^q(y)f_1}, \\
  & \langle \sin(\varphi-2\varphi_{TT}) \rangle_{U,TT} = -x\kappa_M k_{\perp M} \frac{  T_{3}^q(y)g_{TT}}{T_{0 }^q(y)f_1}, \\
  & \langle \cos(3\varphi-3\varphi_{TT}) \rangle_{U,TT} = x\kappa_M k_{\perp M}^3\frac{ T_{2}^q(y)f^\perp_{TT}}{2T_{0 }^q(y)f_1}, \\
  & \langle \sin(3\varphi-2\varphi_{TT}) \rangle_{U,TT} = x\kappa_M k_{\perp M}^3\frac{  T_{3}^q(y)g^\perp_{TT}}{2T_{0 }^q(y)f_1}.
\end{align}
These azimuthal asymmetries can be measured to extract the corresponding distribution functions.

\subsection{The intrinsic asymmetries}

\begin{figure}
\centering
\includegraphics[width= 0.4\linewidth]{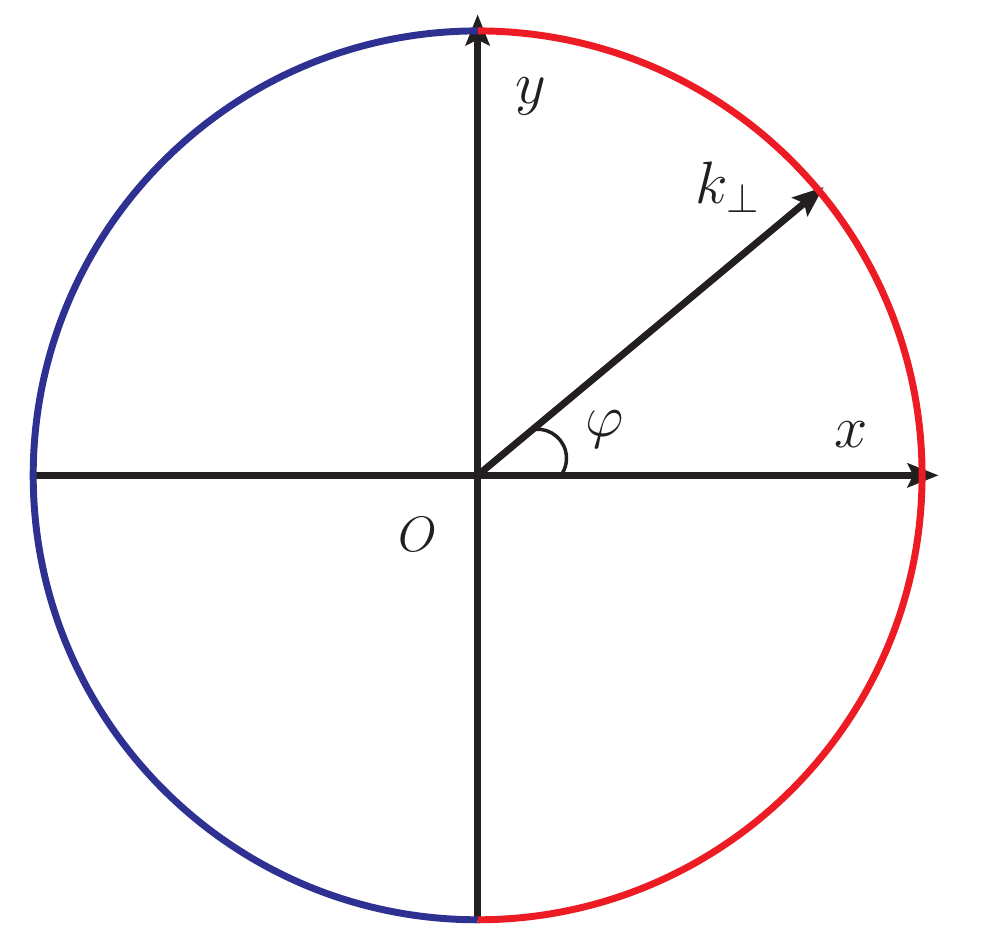}
\caption{The quark (jet) transverse momentum in the $x-y$ plane. The difference of the red hemisphere and the blue one gives the imbalance of the transverse momentum distribution.}
\label{fig:asy}
\end{figure}

As mentioned in the Introduction that the final jet in SIDIS can be a direct probe of analyzing properties of the quark transverse momentum in the $VN$ collinear frame because the transverse momentum of jet is equal to that of the quark. In this part we calculate the intrinsic asymmetry which was first introduced in Ref. \cite{Yang:2022xwy} to explore the transverse momentum distribution of the quark in a nucleon.

In the $VN$ collinear frame, the $z$ direction is defined by the momentum of the nucleon, the transverse momentum of the incident quark (jet) is therefore in the $x-y$ plane. It can be decomposed as follows
\begin{align}
 & k_\perp^{x}=k_\perp \cos\varphi, \label{f:kperpx}\\
 & k_\perp^{y}=k_\perp \sin\varphi. \label{f:kperpy}
\end{align}
Considering Eq. (\ref{f:kperpx}), it is possible to explore the difference of the momentum in the $x$ direction, i.e., $k_\perp^x (+x)-k_\perp^x (-x)$, see Fig. \ref{fig:asy}. We assume that this difference would be induced by the intrinsic transverse momentum of the quark or the radiation of the gluon in the nucleon. In order to explore this difference, we here introduce the definition of  the intrinsic asymmetry:
\begin{align}
 A^x = \frac{\int_{-\pi/2}^{\pi/2} d\varphi ~d\tilde{\sigma} -\int_{\pi/2}^{3\pi/2} d\varphi~ d\tilde{\sigma}}{\int_{-\pi/2}^{3\pi/2} d\tilde{\sigma}_{U,U} d\varphi}. \label{f:akx}
\end{align}
Here $d\tilde \sigma$ is used to denote $\frac{d\sigma}{dx dy d\psi d^2 k_\perp^\prime}$. Subscript $U,U$ denotes the unpolarized cross section. From Eq. (\ref{f:kperpy}), we can define the asymmetry in the $y$ direction in the similar way but with different integral intervals:
\begin{align}
 A^y = \frac{\int_{0}^{\pi} d\varphi ~d\tilde{\sigma} -\int_{-\pi}^{0} d\varphi~ d\tilde{\sigma}}{\int_{-\pi}^{\pi} d\tilde{\sigma}_{U,U} d\varphi}. \label{f:aky}
\end{align}

According to the definition in Eq. (\ref{f:akx}) and (\ref{f:aky}), we calculate these asymmetries which can be written as in the following forms:
\begin{align}
 & A_U^{NC,x} =-\frac{4x \kappa_M k_{\perp M}}{\pi} \frac{T^q_2(y)f^\perp}{T^q_0(y)f_1}, \label{f:aux} \\
 & A_U^{NC,y} =-\frac{4x \kappa_M k_{\perp M}}{\pi} \frac{T^q_3(y)g^\perp}{T^q_0(y)f_1}, \\
 & A_L^{NC,x} =\frac{4x \kappa_M k_{\perp M}}{\pi} \frac{T^q_3(y)g_L^\perp}{T^q_0(y)f_1}, \\
 & A_L^{NC,y} =-\frac{4x \kappa_M k_{\perp M}}{\pi} \frac{T^q_2(y)f_L^\perp}{T^q_0(y)f_1}, \\
 & A_{LL}^{NC,x} =-\frac{4x \kappa_M k_{\perp M}}{\pi} \frac{T^q_2(y)f_{LL}^\perp}{T^q_0(y)f_1}, \\
 & A_{LL}^{NC,y} =-\frac{4x \kappa_M k_{\perp M}}{\pi} \frac{T^q_3(y)g_{LL}^\perp}{T^q_0(y)f_1},
\end{align}
where superscript $NC$ denotes the neutral current.

It is interesting to show all the numerical results of these asymmetries given above. However, not all of them have proper parameterizations. Model calculation made some attempts but there were large uncertainties \cite{Bacchetta:2008af,Mao:2013waa,Yang:2018aue,Liu:2021ype}. We therefore limit ourselves by only presenting the numerical values of $A_{U}^{NC,x}$ in Fig. \ref{fig:Axuncy-u} and Fig. \ref{fig:Axuncy-d}. We take the Gaussian ansatz for the TMD PDFs, i.e.,
\begin{align}
 & f_1(x, k_\perp) =\frac{1}{\pi {\Delta^\prime}^2} f_1(x) e^{-\vec k_\perp^2/{\Delta^\prime}^2},  \label{f:f1xk}
\end{align}
where $f_1(x)$ are taken from CT14 \cite{Schmidt:2015zda} and the faction is taken as $x=0.3$ for illustration. In order to determine $f^\perp(x, k_\perp)$, we have used the Wandzura-Wilczek approximation (neglecting quark-gluon-quark correlation function, $g=0$) \cite{Mulders:1995dh,Bacchetta:2006tn}. It is given by
\begin{align}
 & f^\perp(x, k_\perp) =\frac{1}{\pi \Delta^2x}f_1(x) e^{-\vec k_\perp^2/\Delta^2}.\label{f:fperpxk}
\end{align}
In the numerical estimate, only the up and down quarks are taken into account. The widths of the unpolarized distribution function $f_1(x,k_\perp)$ are taken as ${\Delta^\prime}^2_u={\Delta^\prime}^2_d= 0.53~\rm{GeV}^2 $ \cite{Anselmino:2005nn,Signori:2013mda,Anselmino:2013lza,Cammarota:2020qcw,Bacchetta:2022awv}.
In Fig. \ref{fig:Axuncy-u}, the width of the distribution function $f^\perp(x,k_\perp)$ for up quark is fixed as $\Delta_{u}^2= 0.5~\rm{GeV}^2 $. In Fig. \ref{fig:Axuncy-d}, the width of the distribution function $f^\perp(x,k_\perp)$ for down quark is fixed as $\Delta_{d}^2= 0.5~\rm{GeV}^2 $.  We find that the intrinsic asymmetry $A^{NC,x}_U$ decreases with respect to the energy and it is more sensitive to $\Delta_u^2$ than $\Delta_d^2$.
\begin{widetext}

\begin{figure}
\centering
\includegraphics[width= 0.35\linewidth]{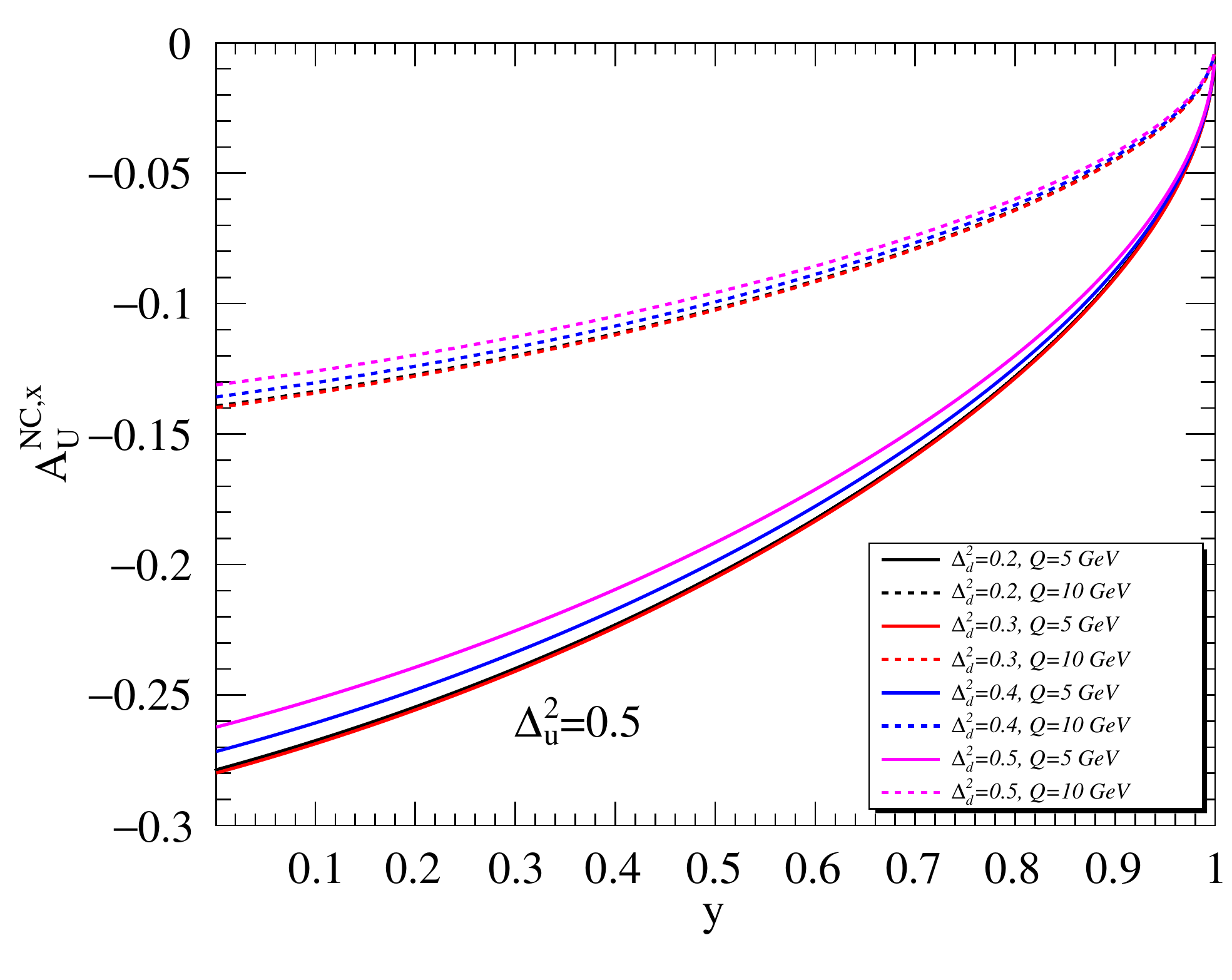}
\quad \quad \quad
\includegraphics[width= 0.35\linewidth]{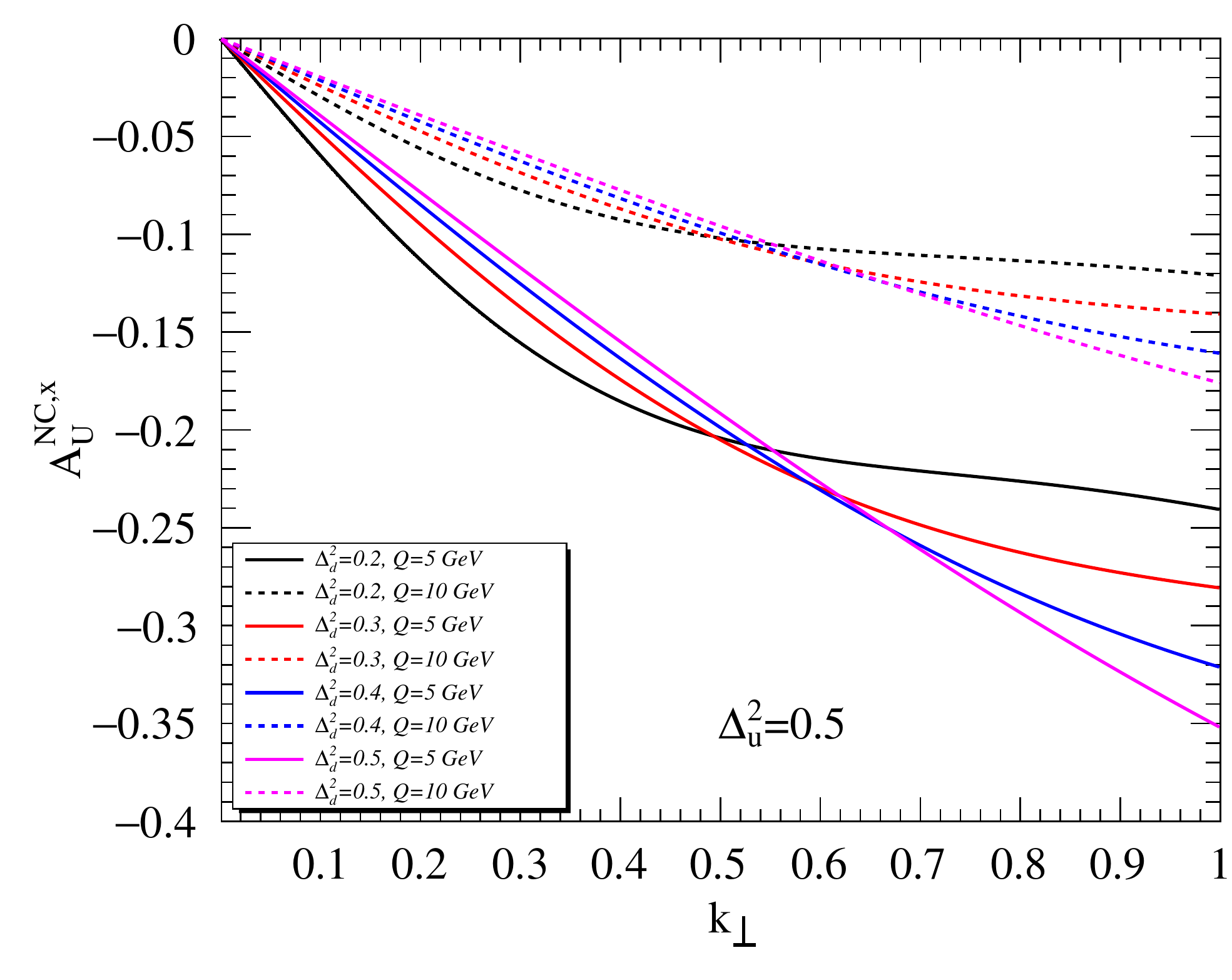}
\caption{The intrinsic asymmetry $A_{U}^{NC,x}$ with respect to $y$ (left) and $k_\perp$ (right). The solid lines show the asymmetry at $Q=$ 5~GeV while the dashed lines show the asymmetry at $Q=$ 10~GeV. Here ${\Delta^\prime}^2_u={\Delta^\prime}^2_d= 0.53$~ and $\Delta_{u}^2= 0.5~\rm{GeV}^2 $, $\Delta_{d}^2$ runs from $0.2$ to $0.5 ~\rm{GeV}^2 $.}
\label{fig:Axuncy-u}
\end{figure}
\begin{figure}
\centering
\includegraphics[width= 0.35\linewidth]{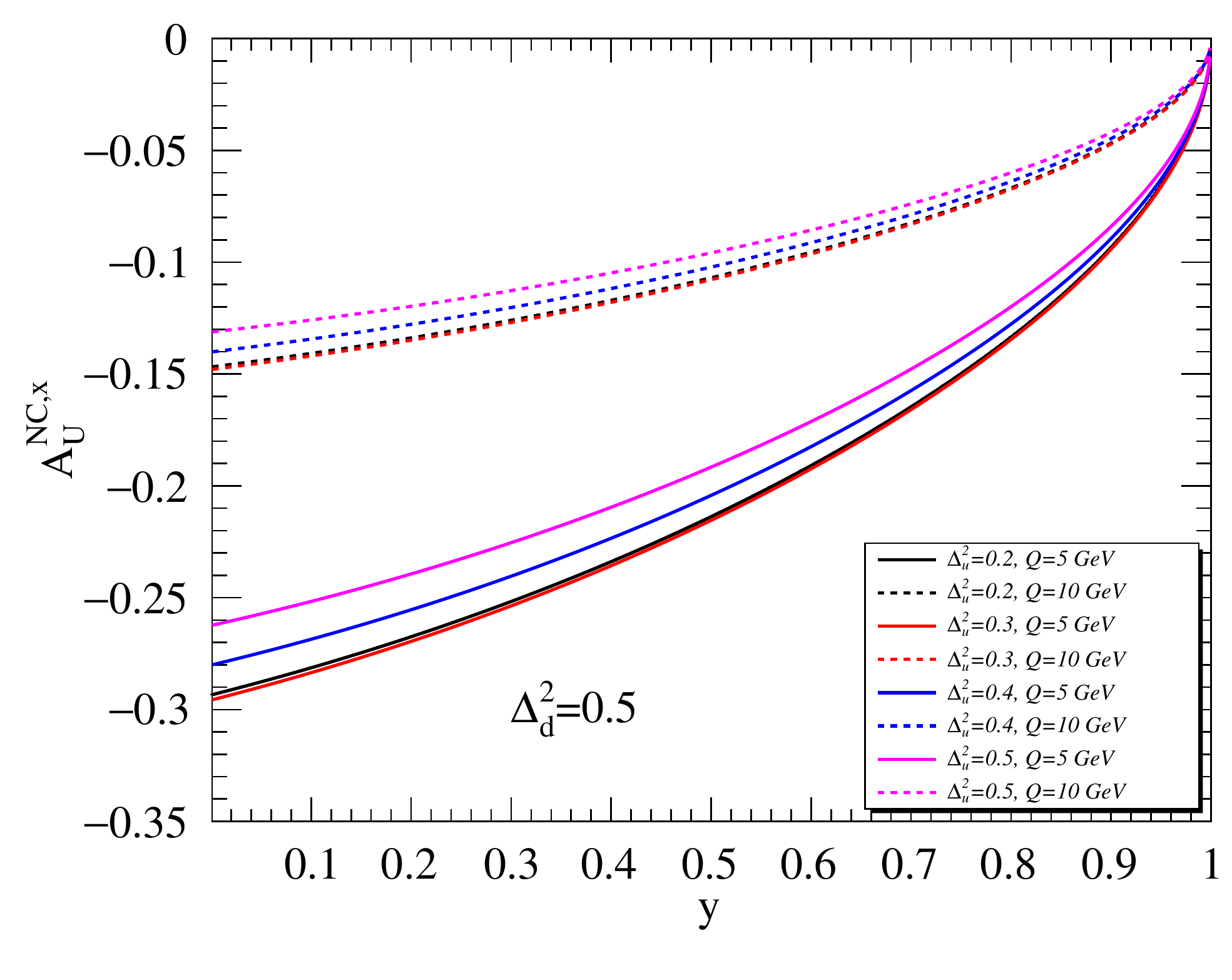}
\quad \quad \quad
\includegraphics[width= 0.35\linewidth]{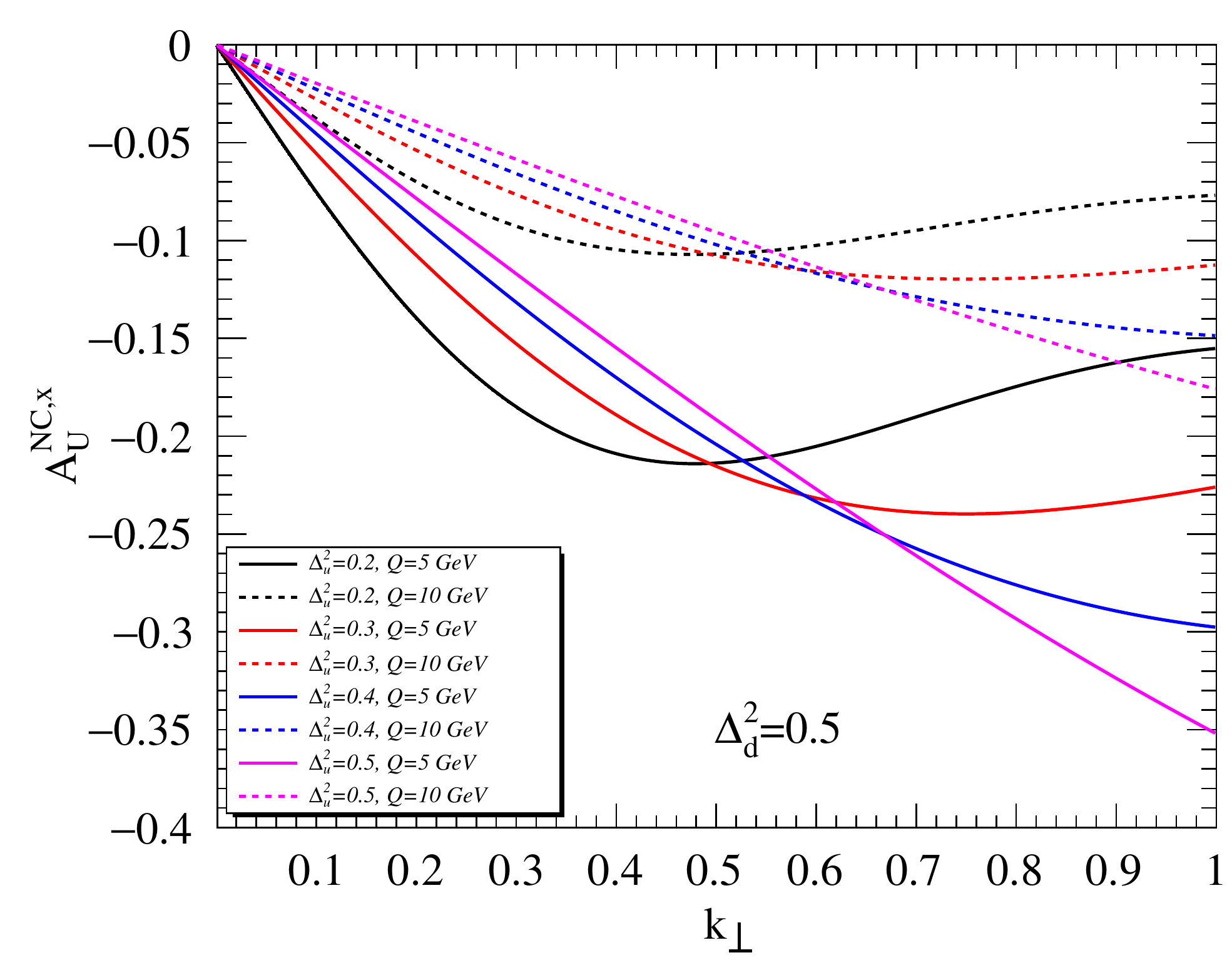}
\caption{The intrinsic asymmetry $A_{U}^{NC,x}$ with respect to $y$ (left) and $k_\perp$ (right). The solid lines show the asymmetry at $Q=$ 5~GeV while the dashed lines show the asymmetry at $Q=$ 10~GeV. Here ${\Delta^\prime}^2_u={\Delta^\prime}^2_d= 0.53$~ and $\Delta_{d}^2= 0.5~\rm{GeV}^2 $, $\Delta_{u}^2$ runs from $0.2$ to $0.5 ~\rm{GeV}^2 $.}
\label{fig:Axuncy-d}
\end{figure}

\end{widetext}

\section{Results of the charged current process}\label{sec:charged}

In this section we present the differential cross section, azimuthal asymmetries and intrinsic asymmetries in the charged current (anti)neutrino nucleus scattering process.

\subsection{The differential cross section}

As before, we divide the differential cross section into a leading twist part and a twist-3 part. Without showing the tedious calculation process, we just give the final result. The leading twist differential cross section is given by
\begin{align}
  \frac{d\sigma_{\nu N,t2}^{CC}}{dx dy d\psi d^2 k_\perp^\prime} =\frac{\alpha_{\rm em}^2 A_{W}}{2y Q^2} T_0(y) &\Biggl\{(f_1+S_{LL}f_{1LL}) - \lambda_h g_{1L} \nonumber\\
  +|S_T|k_{\perp M}&\Big[\sin(\varphi-\varphi_S) f^\perp_{1T}\nonumber\\
  -&\cos(\varphi-\varphi_S) g^\perp_{1T}\Big] \nonumber\\
  -|S_{LT}|k_{\perp M}&\Big[\sin(\varphi-\varphi_{LT}) g^\perp_{1LT}\nonumber\\
  +&\cos(\varphi-\varphi_{LT})  f^\perp_{1LT}\Big] \nonumber\\
  -|S_{TT}|k_{\perp M}^2&\Big[\sin(2\varphi-2\varphi_{TT}) g^\perp_{1TT}\nonumber\\
  - &\cos(2\varphi-2\varphi_{TT})f^\perp_{1TT}\Big]
 \Biggr\},\label{f:crosst2CCnu}
\end{align}
where $T_0(y) = A(y) +  C(y)$. We note that the charged current neutrino nucleus scattering would select the negative charged quark flavors ($d, s, \cdots$) and the CKM matrix element should be inserted in the Eq. (\ref{f:crosst2CCnu}). The twist-3 differential cross section is given by
\begin{align}
  \frac{d\sigma_{\nu N,t3}^{CC}}{dx dy d\psi d^2 k_\perp^\prime} =&- \frac{\alpha_{\rm{em}}^2 A_{W} }{2y Q^2}2x\kappa_M T_2(y)\Biggl\{ \lambda_h k_{\perp M}\nonumber\\
  &\times\Big[\sin\varphi f^\perp_L- \cos\varphi g_L^\perp\Big] \nonumber\\
  &+k_{\perp M}\cos\varphi (f^\perp+S_{LL}f^\perp_{LL}) \nonumber\\
  &+k_{\perp M}\sin\varphi  (g^\perp+S_{LL}g^\perp_{LL}) \nonumber\\
  &+|S_T|\Big[\sin\varphi_S  f_T -\cos\varphi_S g_T  \nonumber\\
  &\quad \quad +\sin(2\varphi-\varphi_S) \frac{k_{\perp M}^2}{2}f^\perp_T \nonumber\\
  &\quad \quad -\cos(2\varphi-\varphi_S) \frac{k_{\perp M}^2}{2}g^\perp_T \Big] \nonumber\\
  &+|S_{LT}|\Big[\sin\varphi_{LT} g_{LT} +\cos\varphi_{LT} f_{LT}  \nonumber\\
  &\quad \quad +\sin(2\varphi-\varphi_{LT})\frac{k_{\perp M}^2}{2} g^\perp_{LT} \nonumber\\
  &\quad \quad +\cos(2\varphi-\varphi_{LT})\frac{k_{\perp M}^2}{2} f^\perp_{LT} \Big] \nonumber\\
  &+|S_{TT}|\Big[\sin(\varphi-2\varphi_{TT}) k_{\perp M} g_{TT} \nonumber\\
  &\quad \quad -\cos(\varphi-2\varphi_{TT}) k_{\perp M} f_{TT} \nonumber\\
  &\quad -\sin(3\varphi-2\varphi_{TT}) \frac{k_{\perp M}^3}{2} g^\perp_{TT} \nonumber\\
  &\quad -\cos(3\varphi-2\varphi_{TT}) \frac{k_{\perp M}^3}{2} f^\perp_{TT} \Big]
 \Biggr\}, \label{f:crosst3CCnu}
\end{align}
where $T_2(y) = B(y) + D(y)$.
Equations (\ref{f:crosst2CCnu}) and (\ref{f:crosst3CCnu}) give the differential cross section of the charged current neutrino nucleus scattering.

To obtain the corresponding antineutrino nucleus scattering, we only need to replace $T_{0,2}(y)$ by  $t_{0,2}(y)$ which are defined as
\begin{align}
  & t_0(y) = A(y) -  C(y), \\
  & t_2(y) = B(y) -  D(y).
\end{align}
We also note here the antineutrino nucleus scattering selects the positive charged quarks ($u, c, \cdots$) and the corresponding CKM matrix element should be inserted in the expression which is neglected here for simplicity.

\subsection{The azimuthal asymmetries}

With the same method given in the previous section, we here show the azimuthal asymmetries in the charged current process. For the leading twist asymmetries, we have
\begin{align}
 & \langle \sin(\varphi-\varphi_S) \rangle_{U,T} = k_{\perp M} \frac{T_{0} (y) f^\perp_{1T}}{2 T_{0} (y) f_1}, \\
 & \langle \cos(\varphi-\varphi_S) \rangle_{U,T} = - k_{\perp M}\frac{ T_{0} (y) g^\perp_{1T}}{2 T_{0} (y) f_1}, \\
 & \langle \sin(\varphi-\varphi_{LT}) \rangle_{U,LT} = -k_{\perp M} \frac{ T_{0} (y) g^\perp_{1LT}}{2 T_{0} (y)f_1}, \\
 & \langle \cos(\varphi-\varphi_{LT}) \rangle_{U,LT} = - k_{\perp M} \frac{ T_{0} (y) f^\perp_{1LT}}{2 T_{0} (y)f_1}, \\
 & \langle \sin(2\varphi-\varphi_{TT}) \rangle_{U,TT} = - k_{\perp M}^2 \frac{ T_{0} (y)g^\perp_{1TT}}{2 T_{0} (y)f_1}, \\
 & \langle \cos(2\varphi-\varphi_{TT}) \rangle_{U,TT} = k_{\perp M}^2 \frac{ T_{0} (y)f^\perp_{1TT}}{2 T_{0} (y)f_1}.
\end{align}
For the twist-3 asymmetries, we have
\begin{align}
  & \langle \cos\varphi \rangle_{U,U} = -x\kappa_M k_{\perp M} \frac{ T_{2} (y)f^\perp}{T_{0} (y)f_1}, \\
  & \langle \sin\varphi \rangle_{U,U} = -x\kappa_M k_{\perp M} \frac{T_{2} (y)g^\perp}{T_{0 } (y)f_1}, \\
  & \langle \cos\varphi \rangle_{U,L} = -x\kappa_M k_{\perp M} \frac{T_{2} (y) f^\perp -\lambda_h T_{2} (y)g^\perp_L}{T_{0} (y) f_1}, \\
  & \langle \sin\varphi \rangle_{U,L} = -x\kappa_M k_{\perp M} \frac{T_{2} (y) g^\perp + \lambda_h T_{2} (y)f^\perp_L}{T_{0} (y)f_1}, \\
  & \langle \cos\varphi \rangle_{U,LL} = -x\kappa_M k_{\perp M} \frac{T_{2} (y)(f^\perp + S_{LL}f^\perp_{LL})}{T_{0} (y)f_1}, \\
  & \langle \sin\varphi \rangle_{U,LL} = -x\kappa_M k_{\perp M} \frac{T_{2} (y) (g^\perp + S_{LL}g^\perp_{LL})}{T_{0} (y)f_1}, \\
  & \langle \cos\varphi_S \rangle_{U,T} =  x\kappa_M\frac{T_{2} (y)g_T}{T_{0 } (y)f_1}, \\
  & \langle \sin\varphi_S \rangle_{U,T} = -x\kappa_M\frac{T_{2} (y)f_T}{T_{0 } (y)f_1}, \\
  & \langle \cos(2\varphi-\varphi_S) \rangle_{U,T} = x\kappa_M k_{\perp M}^2\frac{ T_{3} (y)g^\perp_T}{2T_{0 } (y)f_1}, \\
  & \langle \sin(2\varphi-\varphi_S) \rangle_{U,T} = -x\kappa_M k_{\perp M}^2\frac{T_{2 } (y)f^\perp_T}{2T_{0 } (y)f_1}, \\
  & \langle \cos\varphi_{LT} \rangle_{U,LT} = - x\kappa_M\frac{T_{2} (y)f_{LT}}{T_{0 } (y)f_1}, \\
  & \langle \sin\varphi_{LT} \rangle_{U,LT} = -x\kappa_M\frac{T_{2} (y)g_{LT}}{T_{0 } (y)f_1}, \\
  & \langle \cos(2\varphi-\varphi_{LT}) \rangle_{U,LT} = -x\kappa_M k_{\perp M}^2\frac{ T_{2} (y)f^\perp_{LT}}{2 T_{0 } (y)f_1}, \\
  & \langle \sin(2\varphi-\varphi_{LT}) \rangle_{U,LT} = -x\kappa_M k_{\perp M}^2\frac{ T_{2} (y)g^\perp_{LT}}{2 T_{0 } (y)f_1}, \\
  & \langle \cos(\varphi-2\varphi_{TT}) \rangle_{U,TT} =  x\kappa_M k_{\perp M} \frac{  T_{2} (y)f_{TT}}{T_{0 } (y)f_1}, \\
  & \langle \sin(\varphi-2\varphi_{TT}) \rangle_{U,TT} = -x\kappa_M k_{\perp M} \frac{  T_{2} (y)g_{TT}}{T_{0 } (y)f_1}, \\
  & \langle \cos(3\varphi-3\varphi_{TT}) \rangle_{U,TT} = x\kappa_M k_{\perp M}^3\frac{ T_{2} (y)f^\perp_{TT}}{2T_{0 } (y)f_1}, \\
  & \langle \sin(3\varphi-2\varphi_{TT}) \rangle_{U,TT} = x\kappa_M k_{\perp M}^3\frac{  T_{2} (y)g^\perp_{TT}}{2T_{0 } (y)f_1}.
\end{align}
We have in total 6 leading twist and 18 twist-3 azimuthal asymmetries and they have a one-to-one correspondence with that in the neutral current scattering process. All of them can be measured to extract the distribution functions.

\subsection{The intrinsic asymmetries}

The intrinsic asymmetries in the charged current neutrino nucleus scattering process can also be obtained by using the definition given in Eqs. (\ref{f:akx}) and (\ref{f:aky}). The explicit results are given by
\begin{align}
 & A_U^{CC,x} =-\frac{4x \kappa_M k_{\perp M}}{\pi} \frac{T_2(y)f^\perp}{T_0(y)f_1}, \\
 & A_U^{CC,y} =-\frac{4x \kappa_M k_{\perp M}}{\pi} \frac{T_2(y)g^\perp}{T_0(y)f_1}, \\
 & A_L^{CC,x} =\frac{4x \kappa_M k_{\perp M}}{\pi} \frac{T_2(y)g_L^\perp}{T_0(y)f_1}, \\
 & A_L^{CC,y} =-\frac{4x \kappa_M k_{\perp M}}{\pi} \frac{T_2(y)f_L^\perp}{T_0(y)f_1}, \\
 & A_{LL}^{CC,x} =-\frac{4x \kappa_M k_{\perp M}}{\pi} \frac{T_2(y)f_{LL}^\perp}{T_0(y)f_1}, \\
 & A_{LL}^{CC,y} =-\frac{4x \kappa_M k_{\perp M}}{\pi} \frac{T_2(y)g_{LL}^\perp}{T_0(y)f_1},
\end{align}
where superscript $CC$ denotes the charged current.

\begin{widetext}

\begin{figure}
\centering
\includegraphics[width= 0.35\linewidth]{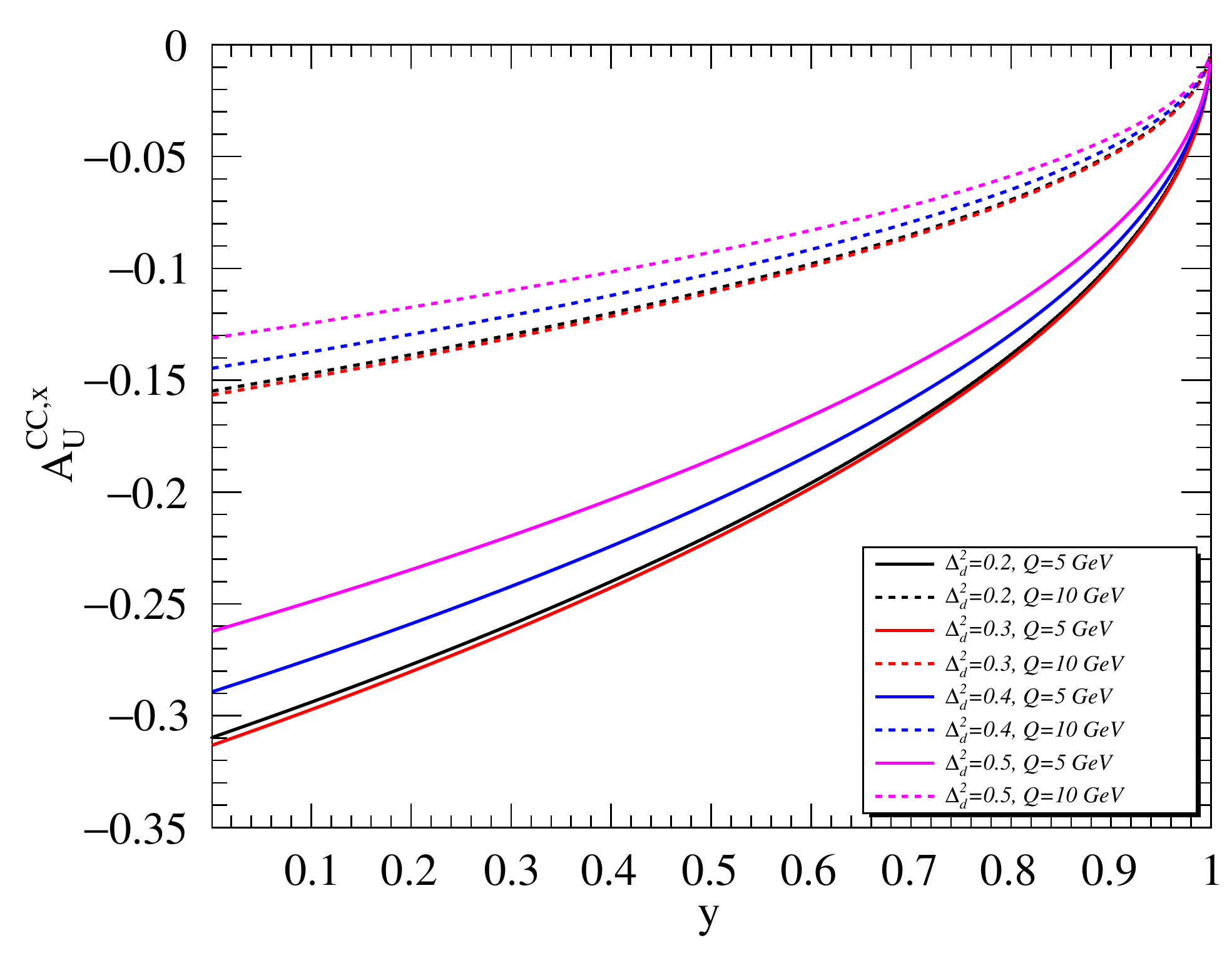}
\quad \quad \quad
\includegraphics[width= 0.35\linewidth]{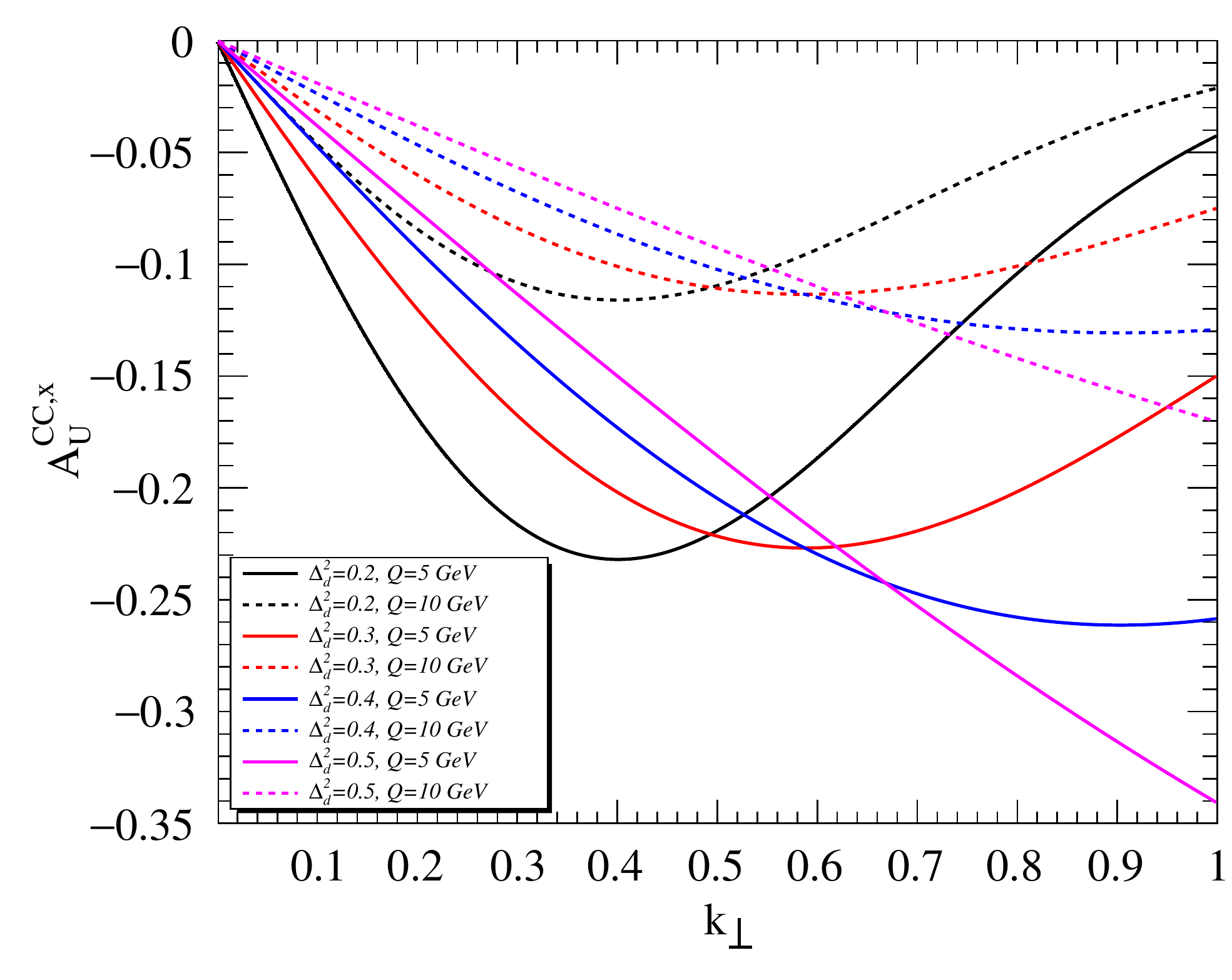}
\caption{The intrinsic asymmetry $A_{U}^{CC,x}$ with respect to $y$ (left) and $k_\perp$ (right). The solid lines show the asymmetry at $Q=$ 5~GeV while the dashed lines show the asymmetry at $Q=$ 10~GeV. Here ${\Delta^\prime}^2_d= 0.53~\rm{GeV}^2 $~ and $\Delta_{d}^2$ runs from $0.2$ to $0.5 ~\rm{GeV}^2 $.}
\label{fig:Axuccy-u}
\end{figure}

\end{widetext}

Using the same parametrizations given in the previous section in Eqs. (\ref{f:f1xk}) and (\ref{f:fperpxk}), we present the numerical values of $A_{U}^{CC,x}$ in Fig. \ref{fig:Axuccy-u}. Since only the down quark is taken into account, the CKM matrix element $U_{ud}$ is approximated as 1. From Fig. \ref{fig:Axuccy-u}, we find that intrinsic asymmetries $A_{U}^{CC,x}$ and $A_U^{NC,x}$ have the same behaviors and the same orders of magnitude. Furthermore, they both decrease with respect to the energy.

\subsection{The plus and minus cross section}

The neutrino and antineutrino nucleus scattering would select different flavors. Then it is helpful to define the $plus$ and $minus$ cross sections. For simplicity, we consider the $k'_\perp$-integrated cross section, i.e., the inclusive cross section.
\begin{align}
  d\sigma^P_{in}=\frac{d\sigma_{in}(\bar\nu)}{dx dy d\psi} + \frac{d\sigma_{in}(\nu)}{dx dy d\psi}, \label{f:crossplus}\\
  d\sigma^M_{in}=\frac{d\sigma_{in}(\bar\nu)}{dx dy d\psi} - \frac{d\sigma_{in}(\nu)}{dx dy d\psi}, \label{f:crossminus}
\end{align}
where subscript $in$ denotes inclusive, $P, M$ denote the plus and minus cross sections calculated by Eqs. (\ref{f:crossplus}) and (\ref{f:crossminus}), respectively. The explicit expressions of the plus and minus cross sections are
\begin{align}
  d\sigma_{in}^P = \frac{\alpha_{\rm{em}}^2 A_{W}}{2y Q^2}\biggl\{& \left[\left(t_0(y)f^U_1(x)+T_0(y)f^D_1(x)\right) \right.\nonumber \\
  + S_{LL}&\left(t_0(y)f^U_{1LL}(x)+T_0(y)f^D_{1LL}(x)\right) \nonumber \\
   -\lambda_h &\left. \left(t_0(y)g^U_{1L}(x)+T_0(y)g^D_{1L}(x)\right)\right] \nonumber\\
  -2x\kappa_M\Big\{|S_T| &\left[\sin\varphi_S \left(t_2(y)f^U_T(x)+T_2(y)f^D_T(x)\right) \right.\nonumber \\
  -&\left. \cos\varphi_S \left(t_2(y)g^U_T(x)+T_2(y)g^D_T(x)\right) \right]  \nonumber\\
  +|S_{LT}|& \left[\sin\varphi_{LT} \left(t_2(y)g^U_{LT}(x)+T_2(y)g^D_{LT}(x) \right) \right. \nonumber\\
  +\cos\varphi_{LT}&\left. \left(t_2(y)f^U_{LT}(x)+T_2(y)f^D_{LT}(x)\right)\right]\Big\}  \biggr\}, \label{f:crossplusin} \\
  d\sigma_{in}^M =  \frac{\alpha_{\rm{em}}^2A_{W} }{2y Q^2}\biggl\{& \left[\left(t_0(y)f^U_1(x)- T_0(y)f^D_1(x)\right) \right.\nonumber \\
  + S_{LL}&\left(t_0(y)f^U_{1LL}(x)- T_0(y)f^D_{1LL}(x)\right) \nonumber \\
   -\lambda_h &\left. \left(t_0(y)g^U_{1L}(x)- T_0(y)g^D_{1L}(x)\right)\right] \nonumber\\
  -2x\kappa_M\Big\{|S_T| &\left[\sin\varphi_S \left(t_2(y)f^U_T(x)- T_2(y)f^D_T(x)\right) \right.\nonumber \\
  -&\left. \cos\varphi_S \left(t_2(y)g^U_T(x)- T_2(y)g^D_T(x)\right) \right]  \nonumber\\
  +|S_{LT}|& \left[\sin\varphi_{LT} \left(t_2(y)g^U_{LT}(x)- T_2(y)g^D_{LT}(x) \right) \right. \nonumber\\
  +\cos\varphi_{LT}&\left. \left(t_2(y)f^U_{LT}(x)- T_2(y)f^D_{LT}(x)\right)\right]\Big\}  \biggr\}. \label{f:crossminusin}
\end{align}
The superscripts $D, U$ denote the $d$-type ($d,s,\cdots$) and $u$-type ($u,c,\cdots$) quarks. The CKM matrix elements are not shown in these expressions.

Assuming the target is longitudinally polarized, we would introduce the spin asymmetry as
\begin{align}
  A_\sigma=\frac{d\sigma^{P/M}_{in,\sigma}(+\sigma)-d\sigma^{P/M}_{in,\sigma}(-\sigma)}{ d\sigma^{P/M}_{in,\sigma}(+\sigma)+d\sigma^{P/M}_{in,\sigma}(-\sigma)}, \label{f:ChargeADef}
\end{align}
where the subscript $\sigma$ denotes the target polarization. Therefore, we have
\begin{align}
  & A_L^P =-\frac{t_0(y)g^U_{1L}(x)+T_0(y)g^D_{1L}(x)}{t_0(y)f^U_1(x)+T_0(y)f^D_1(x)}, \\
  & A_L^M =-\frac{t_0(y)g^U_{1L}(x)-T_0(y)g^D_{1L}(x)}{t_0(y)f^U_1(x)-T_0(y)f^D_1(x)}, \\
  & A_{LL}^P =\frac{t_0(y)f^U_{1LL}(x)+T_0(y)f^D_{1LL}(x)}{t_0(y)f^U_1(x)+T_0(y)f^D_1(x)}, \\
  & A_{LL}^M =\frac{t_0(y)f^U_{1LL}(x)- T_0(y)f^D_{1LL}(x)}{t_0(y)f^U_1(x)-T_0(y)f^D_1(x)}.
\end{align}

Since we have calculated the differential cross sections for both the neutral current and charged current interactions, we can define the Paschos-Wolfenstein ratio \cite{Paschos:1972kj} for the semi-inclusive scattering process:
\begin{align}
  R^-_d =\frac{d\sigma^{NC}_{\nu N}-d\sigma^{NC}_{\bar\nu N}}{d\sigma^{CC}_{\nu N}-d\sigma^{CC}_{\bar\nu N}},
\end{align}
where subscript $d$ denotes the differential cross section. It was first measured by the NuTeV collaboration \cite{NuTeV:2001whx}.
By using the results obtained in the previous section, we have
\begin{align}
  R^-_d =\frac{2A_Z}{A_W} \frac{T^q_0(y)f_1(x)-t^q_0(y)f_1(x)}{T_0(y)f^D_1(x)-t_0(y)f^U_1(x)}. \label{f:Rfu}
\end{align}
At low energy limit, $R^-_d$ in Eq. (\ref{f:Rfu}) can be approximated as
\begin{align}
  R^-_d =\frac{1}{2} \frac{T^q_0(y)f_1(x)-t^q_0(y)f_1(x)}{T_0(y)f^D_1(x)-t_0(y)f^U_1(x)}. \label{f:Rfuapp}
\end{align}
With further approximation, $f_1^u=f_1^d$, we obtain the final result:
\begin{align}
  R^-_d =\frac{1}{2}-\sin^2\theta_W =R^-. \label{f:Rfuiso}
\end{align}
Thus, Eq. (\ref{f:Rfu}) provides an additional method for measuring the weak mixing angle.

\section{Summary} \label{sec:summary}

 In this paper, we calculate the (anti)neutrino induced jet production SIDIS process. Neutrino nucleus scattering processes are important channels in studying nucleon and nucleus structures. They can  provide information on the flavour separation which cannot be realized in the charged lepton (SI)DIS experiments alone.  Our calculations are complete at leading order twist-3 level, whose results are helpful for other topics rely heavily on accurate measurements of the different neutrino scattering processes.
 Basically, we only show the explicit calculation of the neutral current neutrino scattering process. The results for the neutral current antineutrino scattering, the charged current neutrino scattering and the charged current antineutrino scattering can be obtained by replacing the corresponding coefficients. To obtain the results of the antineutrino nucleus scattering for the neutral current reaction, we only need to replace $T^q_{0,1,2,3}(y)$ by $t^q_{0,1,2,3}(y)$. For the charged current reaction, we only need to replace $T_{0,2}(y)$ by $t_{0,2}(y)$ to obtain the results of the antineutrino nucleus scattering.
 Since higher twist effects are significant for semi-inclusive reaction processes and TMD observables, e.g., twist-3 TMD PDFs often lead to azimuthal asymmetries which are different from the leading twist ones. We therefore present the complete results of the azimuthal asymmetries up to twist-3 for the neutrino induced jet production SIDIS. 6 of them are leading twist and 18 of them are twist-3. Considering the equivalence of the transverse momenta of the quark inside the nucleon and the final jet, we calculate the intrinsic asymmetries to explore the imbalance of the  transverse momentum distribution of the quark.  We find that the intrinsic asymmetry decreases with respect to the energy and it is more sensitive to $\Delta_u^2$ than $\Delta_d^2$.  For the charged current (anti)neutrino nucleus scattering process, we further define the $plus$ and $minus$ cross sections to study TMD PDFs.

Previous neutrino heavy nuclear targets inelastic scattering experiments provided fruitful results in determining the weak parameters and parton distributions \cite{CCFR:1997zzq,NuTeV:2001whx,NuTeV:2005wsg,CHORUS:2005cpn}. But a lot of experimental issues need to be settled. For a recent review, see Ref. \cite{NuSTEC:2017hzk}.
The difficulty of measuring these asymmetries presented in this paper is twofold. One is that the cross sections of the neutrino reactions are too small. The other is that jets in the reaction are hard to be determined. For the NC SIDIS, the neutrino and the jet are determined simultaneously. For the CC SIDIS, the charged lepton and jet are determined simultaneously. Overall, the measurement of neutrino experiments requires a high statistics.
In general, inelastic scattering process is much better understood at high energy rather than relative lower energy.
As for the human-made neutrino experiments, the MINER$\nu$A experiment at Fermilab \cite{MINERvA:2016oql} requires medium to high energy (anti)neutrino beams. While the FASER (Forward Search Experiment) program \cite{Feng:2017uoz,FASER:2018bac} and the Forward Physics Facilities (FPFs) \cite{Anchordoqui:2021ghd,Feng:2022inv} at the large hadron collider (LHC) can potentially be used to measure asymmetries due to the high energy of the neutrino beam. FASER program which was proposed to search for new physics has the capability to measure neutrinos at far-forward region downstream from the interaction point. It has reported the first  neutrino interaction candidates at the LHC \cite{FASER:2021mtu}. The FPF is approximately $500-600 $m away from the interaction point and can  measure both the CC and NC (SI)DIS processes.  For the Forward Liquid Argon Experiment, estimates show that the neutrino fluxes can reach $10^{13}$ through a $1\times 1 \rm{m}^2$ cross-sectional area in the range of $10-1000$ GeV \cite{Anchordoqui:2021ghd}. FPF is  also possible to extend the coverage of existing DIS measurements on nuclear targets. Since neutrino induced DIS experiment is an important part of the FPF experiment. These asymmetries presented in this paper can be measured as long as jets can be determined in addition to the scattered leptons.

\section*{ACKNOWLEDGMENTS}
 This work was supported by the Natural Science Foundation of Shandong Province (Grants No. ZR2021QA015 and ZR2021QA040).

\end{document}